\documentclass[preprint, aps]{revtex4}
\usepackage{bm}
\usepackage[colorlinks=true,linkcolor=blue]{hyperref}
\usepackage{amsmath}
\usepackage{amssymb}
\usepackage{amsthm}
\usepackage{amsfonts}
\usepackage{times}
\usepackage{enumerate}
\usepackage{latexsym}
\usepackage{ifpdf}
\usepackage{graphicx}
\usepackage{makeidx}
\expandafter\ifx\csname package@font\endcsname\relax\else
 \expandafter\expandafter
 \expandafter\usepackage
 \expandafter\expandafter
 \expandafter{\csname package@font\endcsname}%
 \fi
\usepackage{color}

\begin{document}

\title{Epitaxial Stabilization of Ultrathin Films of Rare-Earth Nickelates}

\author{D. J. Meyers$^1$,E. J. Moon$^1$, M. Kareev$^1$, I. C. Tung$^2$, B. A. Gray$^1$, Jian Liu$^{1,3}$, M. J. Bedzyk$^2$, J. W. Freeland$^4$, and J. Chakhalian$^1$}
\affiliation{Department of Physics, University of Arkansas, Fayetteville, AR 72701}
\affiliation{Materials Science and Engineering, Northwestern University, Evanston, IL 60208, USA}
\affiliation{Advanced Light Source, Lawrence Berkeley, Berkeley, CA 94720}
\affiliation{Advanced Photon Source, Argonne National Laboratory, Argonne, IL 60439, USA}

\begin{abstract}

We report on the synthesis of ultrathin films of highly distorted EuNiO$_3$ (ENO) grown by interrupted pulse laser epitaxy on YAlO$_3$ (YAO) substrates. Through mapping the phase space of nickelate thin film epitaxy, the optimal growth temperatures were found to scale linearly with the Goldschmidt tolerance factor. Considering the gibbs energy of the expanding film, this empirical trend is discussed in terms of epitaxial stabilization and the escalation of the lattice energy due to lattice distortions and decreasing symmetry. These findings are fundamental to other complex oxide perovskites, and provide a route to the synthesis of other perovskite structures in ultrathin-film form.
\end{abstract}

\maketitle

Transition metal oxides (TMO) have attracted great attention in recent years due to their interesting properties ranging from high temperature superconductivity, exotic magnetism, and the temperature driven metal-insulator transition (MIT). The rare earth nickelates, RNiO$_3$ (R = La...Y), are a group of small charge transfer gap TMOs which all, except for R = La, exhibit such a transition at a temperature designated $T_{MI}$ accompanied by charge order and unusual $E^\prime$- antiferromagnetism\cite{Imada, Scagnoli-mag}.  The  metal-to-insulator transition  has proven to be tunable by epitaxial strain, pressure, carrier doping, and quantum confinement giving promise for future device applications\cite{Liu, Lengsdorf, Scherwitzl,  Tiwari, Cheng, QF}. For Nd and Pr these transitions  occur at the same temperature $T_N=T_{MIT}$, this concurrence has the effect of complicating the interpretation of data obtained for these materials \cite{Zhou}. On the other hand, for smaller rare earth ions ($e.g.$ Eu, Y, etc.) the magnetic transition is separated from the MIT and structural transition by a large temperature gap. 

Owing to the low thermodynamic stability of the nickelates, conventional solid state chemical synthesis requires very high oxygen pressure and temperatures and yields only micron sized single crystals \cite{Torrance, Piamonteze, Lacorre, Alonso-powder}. This has severely limited our understanding of the physics of these interesting compounds. Due to the lack of macroscopic size crystal growth, thin film synthesis is a promising avenue to overcome this obstacle. However, even in the thin film form these materials have thus far proven  difficult to fabricate in a layer-by-layer fashion, becoming arduous upon application of strain\cite{Kaul}. Several recent publications detail the growth of thin film nickelates by metal-organic chemical vapor deposition and sputtering\cite{Lane, Conchon, Aydogdu, Zaghrioui}. Recently thick ($\sim$210 nm) ENO films have been grown by rf magnetron sputtering\cite{Bilewska}, but x-ray diffraction revealed an essentially textured structure of the samples.

In this letter, we present results from the growth of high quality, fully strained ultrathin films (15 unit cells; uc hereafter) of ENO grown on YAO substrates. Growth conditions were varied across the growth regime (temperature, pressure) until the film quality was optimized. A direct comparison to other nickelates from the rare earth family has revealed an unexpected  systematic trend in synthesis parameters required for the layer-by-layer growth, connected to the degree of structural distortion leading to the synthesis of PrNiO$_3$ (PNO) and  YNiO$_3$ (YNO). This phenomenological dependence is elucidated through the model of epitaxial stabilization evolving  with  the lattice energy.

ENO and YNO were grown on YAO (110) and PNO was grown on (LaAlO$_3$)$_{0.3}$-(Sr$_2$AlTaO$_6$)$_{0.7}$ (LSAT) (001) by interrupted pulse laser epitaxy using a KrF excimer laser  ($\lambda=248$ nm) with rapid pulse cycling of 18 Hz; details of this growth mode can be found elsewhere\cite{blank, kareev}. This allows layer-by-layer growth which was  confirmed by the presence of undamped specular intensity oscillations monitored in-situ by our high pressure  reflection high energy electron diffraction (HP-RHEED). The reported samples have lattice mismatches as follows: ENO on YAO -3.2\%, YNO on YAO -2.5\%, and PNO on LSAT 1.4\%. After deposition films were annealed in 1 atm of ultra-pure O$_2$  for 30 minutes. The film quality was then investigated with atomic force microscopy (AFM), X-ray reflectivity (XRR), and resonant X-ray absorption spectroscopy (XAS).  The optimized growth conditions were found to be 610 $^{\circ}$C, P$_{O_2}$ = 150 mTorr with a laser power density between $2.2-2.4$ J/cm$^2$.

Figure 1(a) shows the characteristic time dependent specular intensity. As seen, during the rapid pulse sequence the RHEED intensity drops and then rapidly recovers within a prolonged dwell time, typical of the layer-by-layer growth.  The electron diffraction pattern was taken after annealing to ensure morphological quality, and is shown in Fig. 1(b). The specular, (0 0), and off-specular, (0 1), (0 -1), and half order (indicated by arrows), (0 $\frac{1}{2}$), (0 -$\frac{1}{2}$), Bragg reflections are evident with a streaking pattern characteristic of excellent surface morphology. The expected half-order peaks are due to the structural distortion of the orthorhombic structure\cite{Proffit}. 
AFM imaging yielded an average surface roughness of $<$ 0.7 \AA.

In order to investigate the electronic structure, XAS measurements on the Ni L-edge were obtained in both TEY and TFY modes in the soft X-ray branch at the 4-ID-C beamline at the Advanced Photon Source in Argonne National Labs. The results of the experiment are shown in Fig. 1(b) along with absorption on the reference bulk NNO powder. As seen, the line-shape of Ni L$_2$-edge located at 870.5 eV shows that nickel is in the proper 3+ valence state, confirming that the proper ENO stoichiometry crucial for materials quality was well preserved\cite{Kang}.

In addition, the structural properties were investigated by X-ray scattering in the X-ray Diffraction Facility at Northwestern University using Cu $K\alpha$$_1$ focused radiation with a 4-circle diffractometer; Fig. 1(c) displays the X-ray reflectivity along the (00L) crystal truncation rod obtained in the vicinity of the YAO (220) reflection. The sharp peak at \emph{Q}$_z$ = 3.384 \AA \emph{}$^{-1}$ is the YAO (220) Bragg peak and the broad feature at  \emph{Q}$_z$ = 3.24 \AA \emph{}$^{-1}$ is the (220) peak for the ENO film. The broader Bragg peaks and thickness fringes are from the ENO epitaxial film from which we determined the c-axis lattice constant of ENO to be 3.878 \AA \emph{}; in good agreement with the expected tetragonal expansion from the bi-axial strain. Additionally, thickness fringes testify for the excellent flatness of the samples. The reciprocal space map around the off-specular (103) Bragg intensity for the film and the substrate is shown in the inset of Fig. 1(c). The position of the weak ENO (103) peak relative to the YAO (103) peak further confirms that the ENO epitaxial thin film is coherently strained.

Further insight can be gained from the comparison of the optimal temperatures to the Goldschmidt tolerance factor (\emph{t} = $[r_A-r_O] / [\sqrt{2} (r_B-r_O)]$, where $r_A$ is the rare earth ionic radius, $r_B$ is the transition metal ionic radius, $r_O$ is the oxygen ionic radius) for several members of the rare-earth nickelate family (A = La, Nd, and Eu)\cite{Torrance}. For perovskites, the tolerance factor, \emph{t}, is a measure of structural distortion.  As seen in Fig. 2(a), graphing these three temperatures vs. the tolerance factor revealed a surprisingly linear trend. The linear fit yields the scaling factor  \textit{$\Delta$T/$\Delta$t} $\sim$ 2166 $^{\circ}$C. The cone in Fig. 2(a) represents an approximate range of empirical uncertainty showing the range of growth temperatures that can be varied without completely degrading film quality. To verify the significance of the the scaling dependence, we  synthesized films of PNO based on the tolerance factor, marked as a red circle in Fig. 2(a). A combination of RHEED, AFM, and XRR showed that high quality films were obtained. As a further  test we synthesized one of the most  distorted members of the family, YNO, from a target composed of NiO and Y$_2$O$_3$ precursor materials at the growth temperature \textit{\textit{\emph{\emph{\emph{specifically} predicted}}}} from Fig.2(a) ($\sim$ 565 $^{\circ}$C). The resulting films demonstrate the same high morphological quality as the rest of the rare-earth nickelate family of materials (YNO and PNO data shown in supplemental\cite{supp}).

In order to explain this empirical trend, we consider the model of epitaxial stabilization of thin films\cite{Kaul,Little}. In this phenomenological framework, the relative difference between the energies for the relaxed and epitaxially stabilized phases is given by:
\begin{equation}\small
\label{...}
\Delta G_f''-\Delta G_f'=\Delta G=h[(\Delta g_v''-\Delta g_v')-\frac{\mu}{1-\nu}\epsilon^2] + [\sigma _s''-\sigma _s']
\end{equation}
where $''$ and $'$ indicate the relaxed and epitaxial phases respectively, \emph{h} is the film thickness, $\Delta g_v$ is the Gibb's free energy per unit volume, $\mu$ is the shear modulus, $\nu$ is the Possion's ratio, $\epsilon$ is the lattice mismatch, and $\sigma_s$ is the surface tension. 
Eq. (1) can be further simplified given that (i) all the rare-earth nickelates have the same relaxed chemical phase (i.e. 2RNiO$_3$ $\rightleftharpoons$ 2NiO + R$_2$O$_3$ + 1/2O$_2$) implying  that  $\Delta g_v''$ is practically constant across the series, (ii) so  is $\sigma _s''$ because $\sigma _s''$ $\gg$ $\sigma _s'$, and (iii) the third term is negated by the appropriate lattice mismatch. Under these assumptions Eq. (1) simplifies to: $\Delta G(T,t) \approx A - \Delta g_v'(T,t)$, where A is  constant with respect to temperature and tolerance factor. This behavior is naturally related to the observation that as the tolerance factor is reduced the total energy of the lattice increases because of the reduced symmetry in cation-anion bond positions; the calculated lattice energies, U$_{pot}$, plotted against tolerance factor are shown in Fig. 2(b)\cite{Glasser}; this in turn raises the magnitude of $\Delta  g_v'$\cite{Li}.

To understand the relationship between the growth temperature and the tolerance factor, we plot $\Delta g_v'$ vs. $T$ for several representative films of the family using the linear dependence of the Gibb's energy on $T$ [see Fig. 3]\cite{Greene, footnote}.  As seen, increasing the tolerance factor shifts the phase line lower (due to the lower lattice energy), which in turn causes an increase of the critical temperature $T$* at which the transition from epitaxial to relaxed phases occurs. Each nickelate will then have the specific temperature, $T$*, above which the perovskite phase is chemically unstable (approximated by the top of the cone in Fig. 2(a)). The final consideration is the adatom diffusion $D(T)$; as the temperature is lowered the diffusion is exponentially retarded leading to further increases in surface roughness (approximated by the bottom of the cone, Fig. 2(a))\cite{kareev}. The combination of these factors rationalizes the existence of a set of optimal growth parameters within a progressively narrowing window below $T$*.

To summarize, investigation of the growth temperature as a function of the tolerance factor for several nickelates revealed a surprising linear trend that was used to \textit{predict synthesis conditions} for PNO and YNO films. This scaling behavior was explained in terms of epitaxial stabilization in good agreement with the empirical observations. These findings are certainly not restricted to rare earth nickelates and could be applied to growth of other perovskite-structured families of materials.

JC was funded by grants from majority DOD-ARO (W911NF-11-1-0200) and partially NSF (DMR-0747808). Work at the Advanced Photon Source is supported by the U.S. Department of Energy, Office of Science under grant No. DEAC02-06CH11357. This work made use of the J. B. Cohen X-Ray Diffraction Facility supported by the MRSEC program of the National Science Foundation (DMR-1121262) at the Materials Research Center of Northwestern University.

\newpage

\begin{figure}[h]\vspace{-0pt}
\includegraphics[width=8cm]{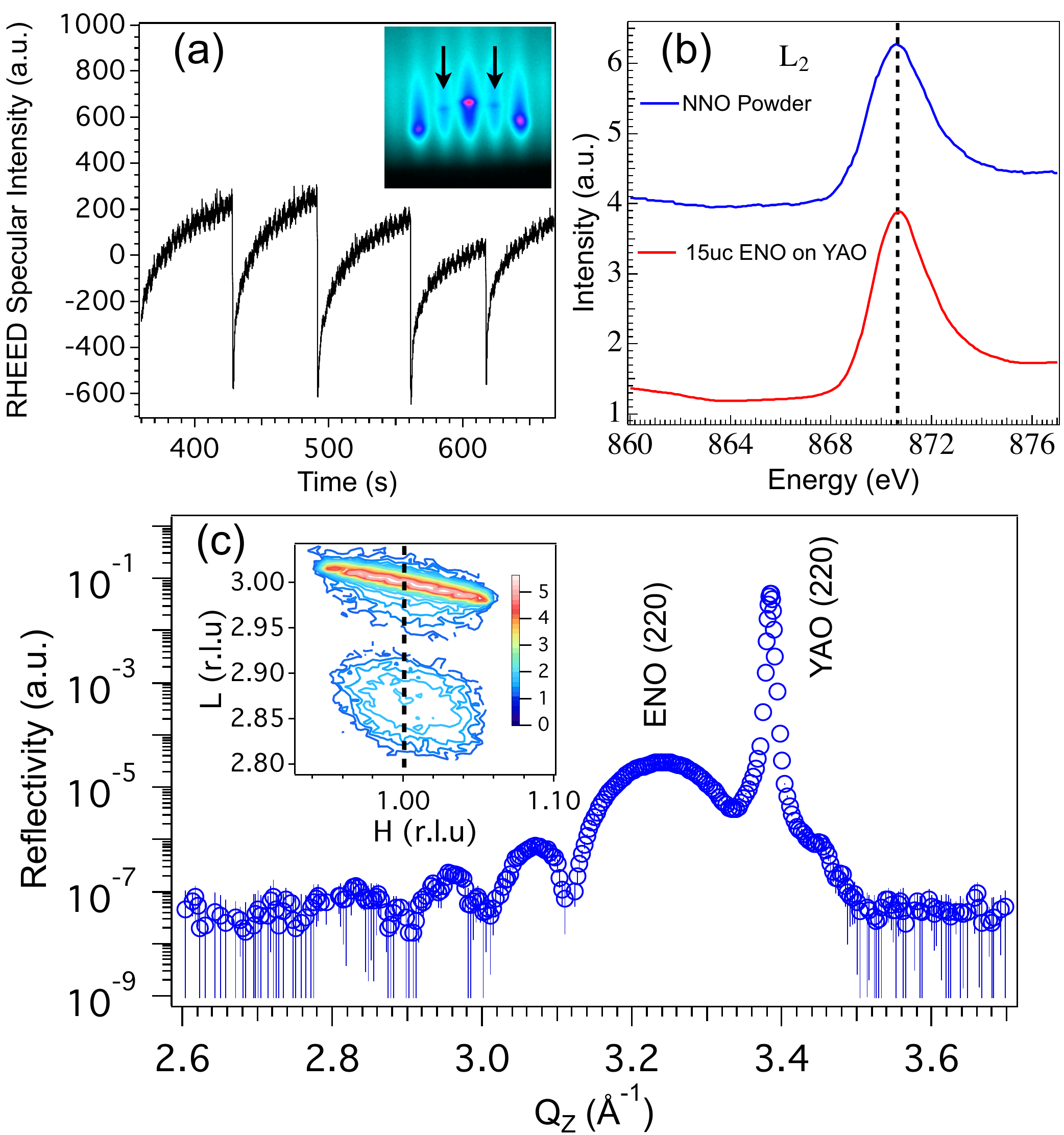}
\caption{\label{} (Color online) (a) RHEED specular intensity taken during growth of 15 uc ENO on YAO. Inset shows the RHEED pattern of the 0th Laue circle on the same ENO sample, black arrows indicate the half-order peaks. (b) Soft x-ray absorption spectroscopy of the Ni L$_2$-edge for a 15uc ENO film (red) and NNO powder (blue). (c) XRD around the (022) direction.The inset shows the RSM at the (103) reflection.}
\end{figure}

\begin{figure}[t]\vspace{-0pt}
\includegraphics[width=8cm]{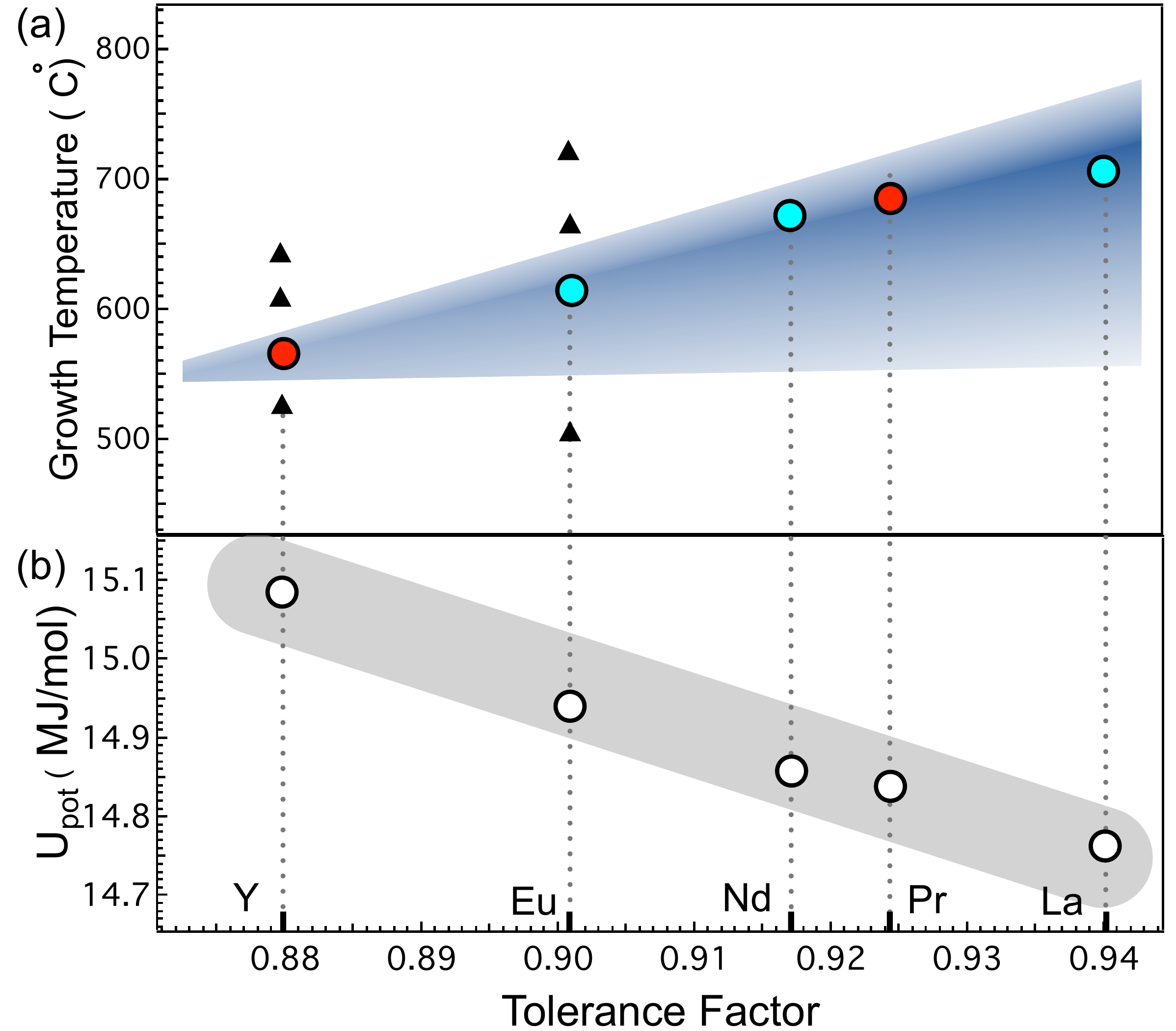}
\caption{\label{} (Color online) (a) Dependence of growth temperature on tolerance factor for nickelates, the dots (green) indicate the data used to obtain the trend, while the red dots were used to synthesize PNO and YNO films. Black triangles signify films grown outside the window (data shown in supplemental\cite{supp}). (b) Lattice energies vs. tolerance factor calculated as described in Ref. 26.}
\end{figure}

 \begin{figure}[t]
\includegraphics[width=8cm]{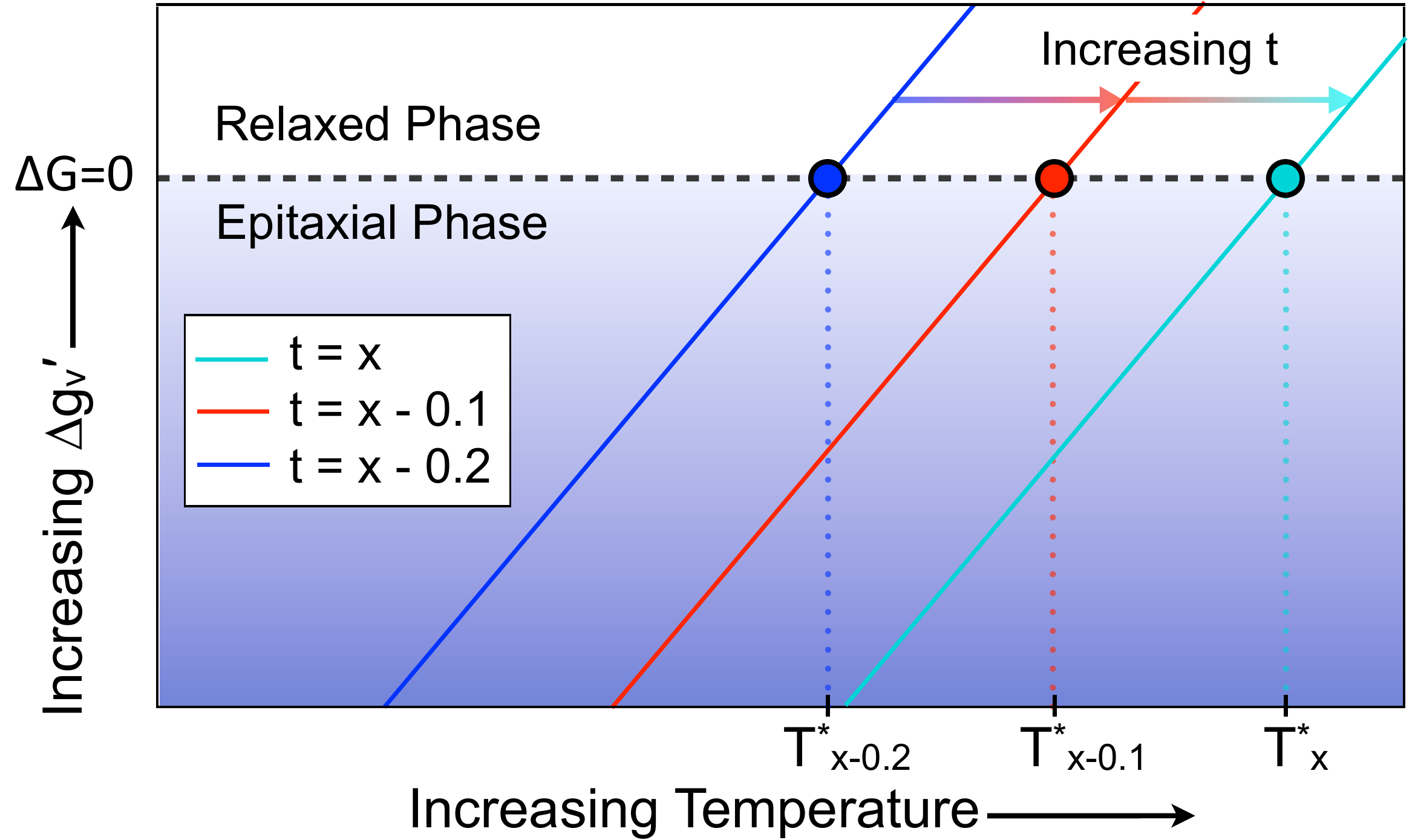}
\caption{\label{} (Color online) Gibb's volume energy versus growth temperature showing the increase in cut-off temperature as tolerance factor is increased.}
\end{figure}


\begin{thebibliography}{999}
\bibitem{footnote}
This also implies that there would be some $T$ dependence for $\Delta g_v''$, but due to the supersaturation being greater for the perovskite phase, it can be disregarded.
\end{thebibliography}
\end{document}